# TEACHING LARGE LANGUAGE MODELS TO SEE IN RADAR: ASPECT-DISTRIBUTED PROTOTYPES FOR FEW-SHOT HRRP ATR[1]

*De Bi[1], Chengbai Xu[1], Lingfeng Chen[1*], Panhe Hu[1]*

[1]*College of Electronic Science, National University of Defense Technology, Changsha 410073, China*
**chenlingfeng@nudt.edu.cn*

## Abstract

High-resolution range profiles (HRRPs) play a critical role in automatic target recognition (ATR) due to their rich information regarding target scattering centers (SCs), which encapsulate the geometric and electromagnetic characteristics of the target. Under few-shot circumstances, traditional learning-based methods often suffer from overfitting and struggle to generalize effectively. The recently proposed HRRPLLM, which leverages the in-context learning (ICL) capabilities of large language models (LLMs) for one-shot HRRP ATR, is limited in few-shot scenarios. This limitation arises because it primarily utilizes the distribution of SCs for recognition while neglecting the variance of the samples caused by aspect sensitivity. This paper proposes a straightforward yet effective Aspect-Distributed Prototype (ADP) strategy for LLM-based ATR under few-shot conditions to enhance aspect robustness. Experiments conducted on both simulated and measured aircraft electromagnetic datasets demonstrate that the proposed method significantly outperforms current benchmarks.

## 1 Introduction

High-Resolution Range Profiles (HRRPs) , *i.e.* 1-D radar line-of-sight (LOS) projections of a target scattering characteristics can be used as effective structural signatures for the critical task of Automatic Target Recognition (ATR) [1, 3, 5, 6, 10, 12, 13, 14, 15, 16, 17, 18]. In recent the past decade, deep learning (DL) has demonstrated marked superiority over traditional methods in the field of HRRP ATR [3, 5, 18]. However, the success of DL is based on the availability of large and comprehensive datasets. This dependency limits real-world applications, especially in non-cooperative circumstances where acquiring labelled data for novel target is often infeasible. Consequently, these models struggle to generalize effectively in few-shot situations [1, 3, 16]. To address this issue, researchers have explored various directions which can be categorized as data augmentation, model simplification, few-shot learning (FSL) [16].

FSL is designed to mimic the human ability to learn and recognize new concepts from a minimal number of examples [2]. Currently, FSL frameworks applied in HRRP ATR include meta-learning [1] and metric-learning [19]. However, current FSL methods exhibit significant limitations: **(a)** These methods often exhibit limited generalization capabilities when the gap between training distribution and testing distribution is big (*i.e.*, limited OOD generalizability) [3, 4]. **(b)** Many of FSL approaches, particularly meta-learning, impose significant computational burdens during the meta-training phase as well as the fine-tuning stage during meta-testing [5]. **(c)** The decision-making process of these black-box models remains opaque, presenting a persistent challenge for explainability [6], which is paramount in high-stakes applications like defense and aerospace.

Recently, the advent of Large Language Models (LLMs) has drawn the attention of the academia. LLMs have demonstrated extraordinary FSL capabilities through its strong in-context learning (ICL) emerged in scaling, where the model learns to perform a task simply by a few examples provided in the prompt, which is training-free [7, 8, 9]. While foundation models are being actively developed for vision [20] and remote sensing domains like SAR imagery [11, 21], the HRRP field has remained largely untouched. This gap is primarily due to two obstacles: the chronic lack of large, publicly available HRRP datasets [10] and, more fundamentally, the modality gap between the electromagnetic nature of HRRP signals and the semantic, text-based domain of LLMs [11]. To solve this, Chen *et al.* [12] proposed HRRPLLM as a framework for HRRP ATR via LLMs based on Scattering Center (SC) textualization. Their results show surprising competitive performance against FSL and traditional baselines. However, HRRPLLM only performed well on one-shot experiments and in the condition where the aspect variance is small. In this paper, we argue that HRRPLLM may induce confusion of aspects under FSL conditions, thereby causing the accuracy of HRRP target recognition to decrease rather than increase when the number of support samples is greater. The fundamental problem is that the general purpose LLMs do not have the knowledge of each target in all aspects, and their ATR is basically a pattern matching that treats all support samples of a class as evidence for a single, monolithic concept. Hence, a mechanism is needed to be designed to help LLMs get the idea fundamentally to ignore the rich, structured intra-class variance induced by aspect sensitivity. This paper proposes the Aspect-Distributed Prototype (ADP) strategy, a simple framework designed to unlock robust few-shot HRRP ATR by explicitly embracing aspect variance within the LLM's ICL process. This strategy constructs a set of distributed prototypes, where each prototype represents a distinct, aspect-dependent signature of the target. This allows the LLM to reason over a distribution of representations, matching a query sample to the most relevant aspect-prototype rather than a implicit single, ambiguous class average.

[1] This paper is a preprint of a paper submitted to the IET International Radar Conference (IRC 2025) and is subject to Institution of Engineering and Technology Copyright. If accepted, the copy of recordwill be available at IET Digital Library.

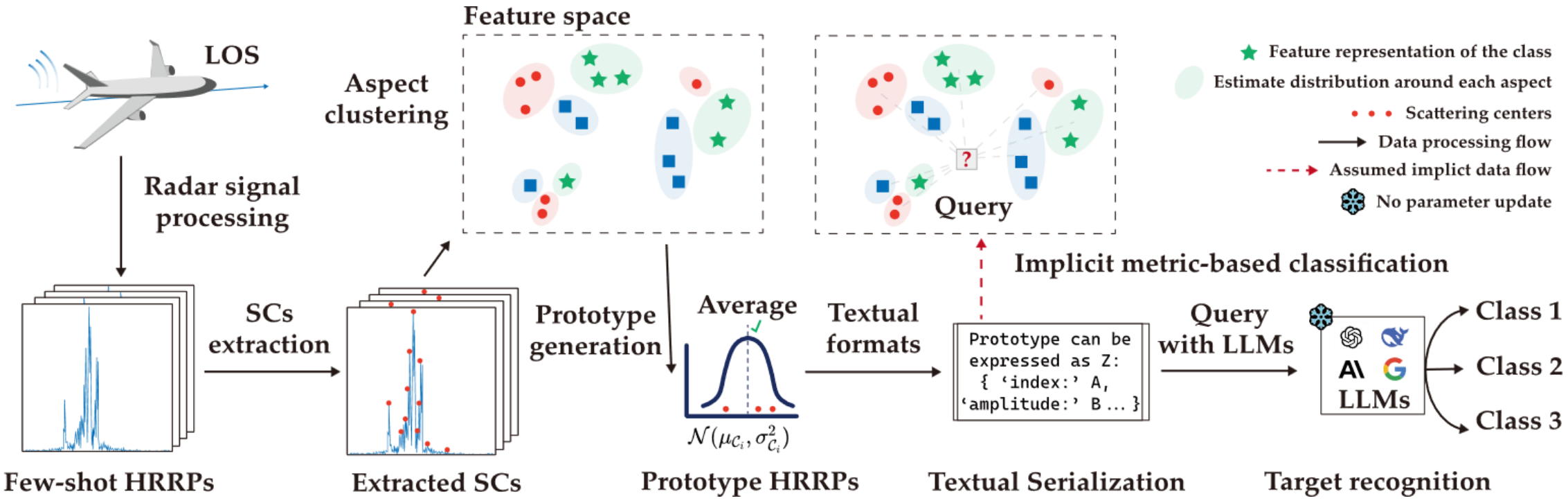

Figure 1 The workflow of the proposed ADP strategy.

**Our contributions. (a)** We are the first to analyze the critical failure mode of LLM-based few-shot HRRP ATR, demonstrating that ignoring aspect-induced variance leads to performance degradation as support samples increase. **(b)** We propose the novel ADP strategy, which explicitly models the intra-class variance of HRRPs. ADP constructs a set of aspect-aware prototypes for each class, enabling the LLM to perform more robust and fine-grained recognition via ICL. **(c)** We conduct experiments on both simulated and measured aircraft electromagnetic datasets. Results show that the proposed method outperforms benchmarks.

## 2 Methodology

### *2.1 Preliminaries*

In few-shot HRRP ATR, we define an $N$-way $K$-shot classification task over $N_0$ query samples as $\mathcal{T}$. This means $N$ distinct target classes and $K$ labelled support shots for each class are sampled to constitute the support set $\mathcal{D}_S = \left\{\left(\mathbf{X}_i^{(S)},\ y_i^{(S)}\right)\right\}_{i=1}^{N\times K}$, where $\mathbf{X}_i^{(S)}$ denotes the $i^{th}$ HRRP support sample and $y_i^{(S)}$ represents its corresponding class label. The query set consists of $N_0$ distinct HRRP samples which can be expressed as $\mathcal{D}_Q = \left\{\left(\mathbf{X}_j^{(Q)},\ y_j^{(Q)}\right)\right\}_{j=1}^{N_0}$ [1, 2]. The model defined by parameters $\Theta$ aims to predict labels $y_j^{(Q)}$ for query samples $\mathbf{X}_j^{(Q)}$ by leveraging the support set $\mathcal{D}_S$, effectively learning a mapping function $f_\Theta(\mathcal{T})\colon \mathbf{X}^{(Q)} \rightarrow \hat{y}^{(Q)}$ where $\hat{y}^{(Q)}$ denotes the predicted label.

### *2.2 HRRP signal formation*

The range resolution $\Delta R$ of HRRP, a key parameter, is calculated by $R = c/(2B)$, where $B$ represents the signal bandwidth, and $c = 3\times 10^8 m/s$ is the speed of light. A common waveform for HRRP imaging is the Linear Frequency Modulated (LFM) signal, whose instantaneous frequency $f(t)$ varies linearly with time: $f(t) = f_0 + Kt$, where $f_0$ is the carrier frequency and $K = B/T_p$ is the chirp rate, in which $T_p$ is the pulse duration. Here, we consider the LFM transmitted signal $S_t(t)$ and denote as:

$$
\begin{aligned}
S_t(t) &= rect\left(\frac{t}{T_p}\right)\exp\left(j2\pi f_0 t\right)\exp\left(j\pi K t^2\right) \\
&= E(t)\exp\left(j2\pi f_0 t\right)
\end{aligned}
\tag{1}
$$

where $rect(t/T_p)$ is the rectangular pulse of width $T_p$, $E(t)$ represents the envelope of the transmitted signal $S_t(t)$.

A HRRP sequence can be taken as a sum of echoes from target dominant SCs along direction of radar detection. In a wideband radar system, the HRRP of target consists of many range cells, and the target echo signal within a range cell can be regarded as the coherent superposition of echo signals from each SC in that range cell. After dechirping processing, the received target echo signal of the $n^{th}$ range cell in baseband can be expressed as:

$$
\begin{aligned}
\hat{S}_{r,n}(t) &= \sum_{i=1}^{M} A(i)E\left(t-\frac{2R_i}{c}\right)\exp\left(-j2\pi f_0\frac{2R_i}{c}\right) \\
&= \sum_{i=1}^{M} A(i)E\left(t-\frac{2R_i}{c}\right) \\
&\quad \times\exp\left(-j2\pi f_0\frac{2R_i}{c}+j\pi K\left(t-\frac{2R_i}{c}\right)^2\right)
\end{aligned}
\tag{2}
$$

where $i = 1, 2, \cdots, M$ represents the $i^{th}$ SC in the $n^{th}$ range cell, $n = 1, 2, \cdots, N$, $N$ represents the total number of range cells, $\hat{S}_{r,n}(t)$ represents the $n^{th}$ range cell in received echo signal, and $A(i)$ is the corresponding scattering coefficient.

After Fourier transformation and pulse compression processing, the frequency response is expanded as:

$$
\begin{aligned}
\rho(\omega) &= \mathcal{F}\left(\hat{S}_{r,n}(t)\right) * E^*(\omega) \\
&= \sum_{i=1}^{M} A(i)P(\omega)\exp\left(-2j\pi(\omega+f_0)\frac{2R_i}{c}\right)
\end{aligned}
\tag{3}
$$

$$\rho(k)=\sum_{i=1}^{M}\Omega(i)rect\left(\frac{\omega}{B}\right)\exp\left(-j2\pi\frac{2R_i}{c}(i-1)\Delta\omega\right) \tag{4}$$

where $\rho(k)$ is the $k^{th}$ frequency response, $k\in\{1,2,\cdots,K\}$, $K$ represents the total number of frequency components and $\Delta\omega$ is the frequency interval. Assume that $R_i$ is an integer multiple of range resolution $\Delta R$, then equation (4) can be rewritten as:

$$\rho(k)=\sum_{n=1}^{N}\Omega(n)rect\left(\frac{\omega}{B}\right)\exp\left(-j2\pi\frac{2R_n}{c}(n-1)\Delta\omega\right) \tag{5}$$

where $R_n=n\cdot\Delta R$ represents the locations in $N$ range cells. Based on equation (5), when considering multiple HRRP signals, it can be transformed into the following vector-matrix form. Let:

$$\begin{aligned}\boldsymbol{\phi}(R_n)=\Big[1,&\exp\left(-j\frac{4\pi}{c}R_n\Delta\omega\right),\exp\left(-j\frac{4\pi}{c}R_n2\Delta\omega\right),\\&\cdots,\exp\left(-j\frac{4\pi}{c}R_n(N-1)\Delta\omega\right)\Big]^T\end{aligned} \tag{6}$$

$$\boldsymbol{\Phi}=[\boldsymbol{\phi}(R_1),\boldsymbol{\phi}(R_2),\cdots,\boldsymbol{\phi}(R_N)]$$

Considering the presence of noise in practical applications, $\rho(k)$ can ultimately be expressed as:

$$\boldsymbol{\rho}=\boldsymbol{\Phi}\boldsymbol{\Omega}+\boldsymbol{\eta} \tag{7}$$

where $\boldsymbol{\Phi}$ is the Fourier basis, $\boldsymbol{\Omega}=[\Omega(1),\Omega(2),\cdots,\Omega(N)]^T$ represents the vector of scattering coefficient and $\boldsymbol{\eta}=[\eta(1),\eta(2),\cdots,\eta(N)]^T$ denotes the noise component. According to equation (7), HRRP can be approximated by dominant SCs [16, 17]. Additionally, the scattering coefficient $\Omega$ contains locations and intensity of SCs.

### 2.3 Aspect-distributed prototypes

Following HRRPLLM's method [7, 12], we perform SCs' extraction on HRRP samples and represent them as a sequence of location and amplitude $\mathcal{S}_{\mathbf{X}}=[p_1,p_2,\cdots,p_j,a_1,a_2,\cdots,a_j]$, where $p_j$ and $a_j$ are the position and amplitude of the $j^{th}$ SC. Individual HRRP sample is highly sensitive to minor variations in target aspect. This sensitivity is fundamentally the variance of target SCs. To overcome this, we seek to design a robust few-shot representation for each target class based on aspect-dependent clusters.

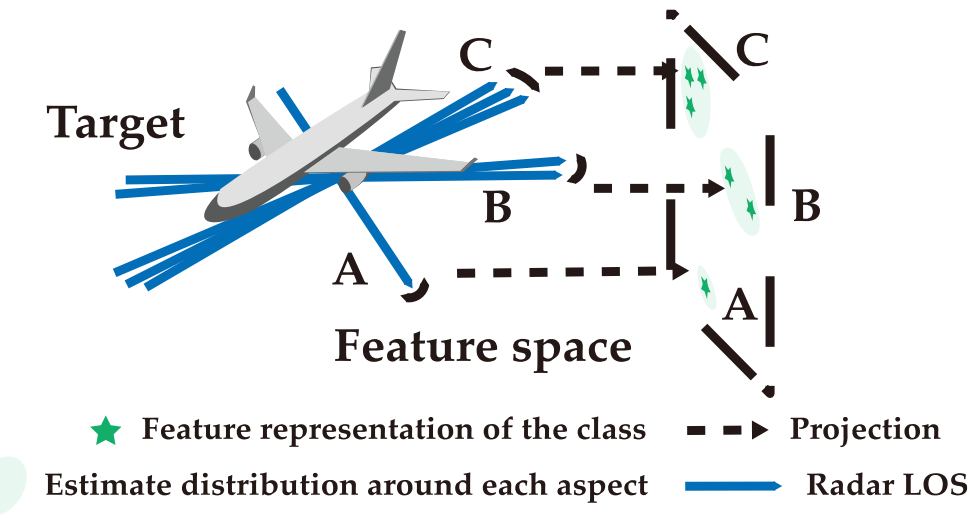

Figure 2 Schematic illustration of the distributed prototype. Due to aspect sessility, distribution of different aspect from different target is often mixed. Since ICL-based method does not adjust feature mapping network. We design prototypical representation with ADP.

Our ADP strategy supposes that when the variance of aspect is ignorable, the distribution of target HRRP follows a Gaussian distribution $n\sim\mathcal{N}(0,\sigma^2)$, where $\sigma^2$ represents variance. Based on this assumption, we first obtained the cluster centers for each class of support set $\mathcal{D}_S$, thereby the samples from the same cluster are supposed to belong to a same aspect. After obtaining the cluster centers, we calculate the averaged prototype in every aspect cluster of each class. As shown in Figure 2, through this way a distributed set of aspect prototypes is obtained for each class, comprehensively representing the distribution of target SCs in all aspects.

### 2.4 Few-shot recognition via ICL

Our method leverages the powerful ICL capability of LLMs. LLMs are conditioned to perform a few-shot task $\mathcal{T}$ through a structured prompt $\mathcal{P}$, requiring no updates to LLM's parameter $\Theta_M$. The core of the ICL strategy lies in the careful construction of a prompt $\mathcal{P}_T$ that presents the complete N-way K-shot task to the LLM in a single pass. The prompt is meticulously assembled in four parts to guide LLMs:

```
**LLM Prompt Input for HRRP ATR (GPT-o4-mini)

**Task definition: You are a radar target recognition expert, skilled at identifying target types by a
nalyzing their scattering center characteristics.Your task is to accurately identify the target from a
list of candidates based on the provided primary scattering center information (position index and rel
ative amplitude).

**Scattering Center Characteristics Overview: Scattering centers are the primary regions on a target w
here radar echo energy is concentrated. They typically correspond to geometric discontinuities, edges,
corners, or strong reflective surfaces of the target.By analyzing the number of scattering centers, th
eir relative positions (range bin indices), and their respective relative amplitudes, one can infer th
e target's size, shape, and structural features.In this task, the provided scattering center informati
on is extracted from a 1D High-Resolution Range Profile (HRRP) of length 1000 and is sorted in descend
ing order of amplitude.The 'position index' starts counting from 0, representing the position in the o
riginal HRRP sequence. The 'relative amplitude' is normalized (e.g., the maximum value is 1).

**Current Dataset and Task:
The data currently being analyzed originates from **measured HRRP scattering center data**. Candidate
target classes include: `an26, citation, yar42`.

**Reasoning Steps and Requirements:
1.  **Examine Test Sample Scattering Centers**: Carefully observe the data provided in the 'Test Sampl
e Scattering Centers' section. Focus on:
    *   The number of detected scattering centers.
    *   The position indices and relative amplitudes of the strongest few scattering centers.
    *   The approximate distribution pattern of scattering centers across the entire target length (0
to 999) (e.g., concentrated at the front, rear, evenly distributed, etc.).
2.  **Reference Neighboring/Support Samples (if provided)**: Compare the scattering center features of
 the test sample with those of known class samples in the 'Neighboring Training Sample Reference'.
    *   Note the known class of each reference sample and compare the similarity of its scattering cen
ter pattern to the test sample.
3.  **Make a Comprehensive Judgment**: Based on your understanding of scattering center distribution p
atterns for different target types and the comparison with reference samples, determine which candidat
e class the test sample most closely matches.
4.  **Output Format**:
    *   On the first line of your response, please clearly state the predicted target class in the for
mat: `Predicted Target Class: [Fill in one of the candidate class names here]`
    *   In subsequent lines, briefly state your main reasons for this judgment, e.g., based on the num
ber, position, or specific pattern of scattering centers, or similarity to a reference sample.

**Neighboring Training Sample Reference (Support Set Protypes):

--- Reference Prototype 1 ---
Known Target Class: `an26`
Its primary scattering center information:
[
  {'range index': 247, 'normalized amplitude': 1.000},
  {'range index': 236, 'normalized amplitude': 0.629},
  {'range index': 242, 'normalized amplitude': 0.330},
  {'range index': 215, 'normalized amplitude': 0.263},
  {'range index': 207, 'normalized amplitude': 0.224}
]

--- Reference Prototype 2 ---
Known Target Class: `citation`
Its primary scattering center information:
[
  {'range index': 234, 'normalized amplitude': 1.000},
  {'range index': 218, 'normalized amplitude': 0.408},
  {'range index': 250, 'normalized amplitude': 0.253},
  {'range index': 199, 'normalized amplitude': 0.187}
]

--- Reference Prototype 3 ---
Known Target Class: `yar42`
Its primary scattering center information:
[
  {'range index': 242, 'normalized amplitude': 1.000},
  {'range index': 255, 'normalized amplitude': 0.725},
  {'range index': 234, 'normalized amplitude': 0.644},
  {'range index': 250, 'normalized amplitude': 0.499},
  {'range index': 227, 'normalized amplitude': 0.372},
  {'range index': 266, 'normalized amplitude': 0.308},
  {'range index': 213, 'normalized amplitude': 0.264}
]

**Test Sample Scattering Centers (Please predict based on this):**
[
  {'range index': 229, 'normalized amplitude': 1.000},
  {'range index': 250, 'normalized amplitude': 0.931},
  {'range index': 222, 'normalized amplitude': 0.872},
  {'range index': 243, 'normalized amplitude': 0.796},
  {'range index': 235, 'normalized amplitude': 0.346},
  {'range index': 209, 'normalized amplitude': 0.306},
  {'range index': 202, 'normalized amplitude': 0.254},
  {'range index': 189, 'normalized amplitude': 0.215}
]

Please strictly follow the output format requirements.
Predicted Target Class:
```

```
**LLM Output Recognition Results (GPT-o4-mini)

**Predicted Target Class: yark-42.

**Reasons: 1. Number of prominent scattering centers (8) closely matches the relatively rich-center pro-
file of the yark-42 prototype (7 centers), versus only  4 for citation or 5 for an26.
           2. The strongest test peaks at range indices 229, 243 and 250 correspond well to yark-42's key
centers at 242, 255, 234 and 250, both in position and high relative amplitude.
           3. The mid-level peaks around 222 and 235 further mirror yar42's secondary centers (234, 227,
213), whereas an26 and citation lack such a similarly distributed cluster.
```

Figure 2 The prompt fed to LLMs following our ADP strategy for a single cluster per class circumstance.

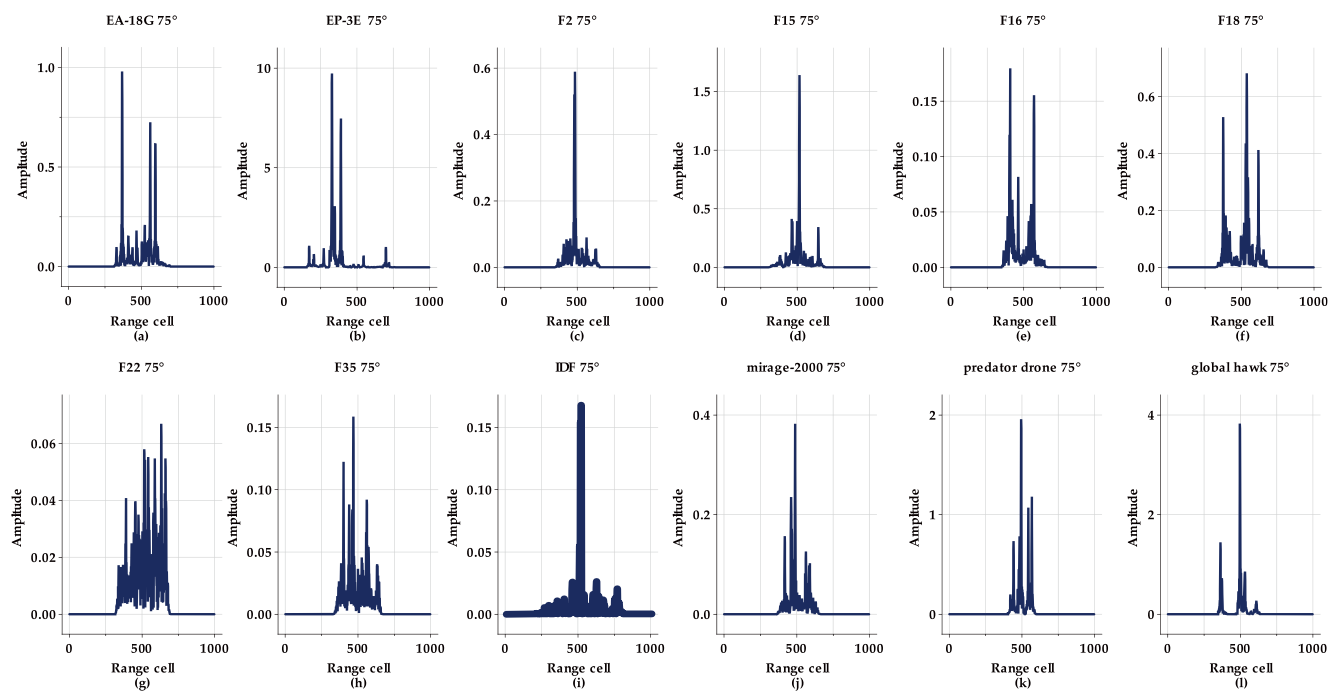

Figure 3 Examples of all 12 types in the simulated dataset at the same aspect.

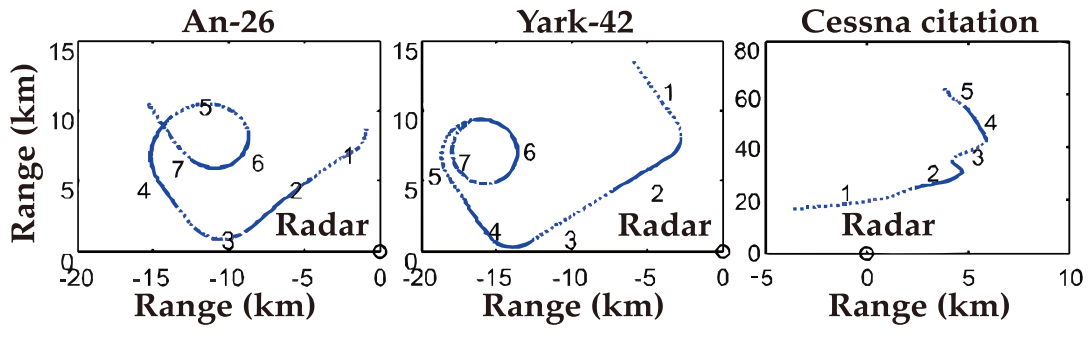

Figure 4 The target trajectories of our 3 types measured dataset.

**i) Domain background information:** This includes a definition of the ATR task, descriptions of HRRPs, SCs and cluster centers, a list of $N$ candidate target classes.

**ii) Reasoning steps:** Several key steps are offered as primary instruction in order to enable LLMs to conduct reasoning aligned with our intended focus points.

**iii) Output format:** It prescribes specific formats for LLMs' responses, including recognition results and explanations.

**iv) Cluster centers of few-shot support set and the query:** support set $\mathcal{D}_S$ consists of $N_1$ cluster centers pairs, where each pair comprises the textual SCs' representation of the cluster center and its true class. Following this, the textual SCs' representation of the query is provided, along with a structured prompt $\mathcal{P}_T$ directing the LLM to predict its class from $N$ candidate classes.

## 3 Experiments

*3.1 Experimental setup*

*3.1.1 Datasets and SC Representation:* **(a)** Measured dataset: this dataset is composed of three aircraft types (An-62, Yark-42, Cessna Citation). The *C*-band radar measurements (central frequency at 5.52 GHz, bandwidth of 400 MHz) generate HRRPs featuring 306 range cells. **(b)** Simulated dataset: 12 similar aircraft classes (*e.g.*, F18, F15, EP-3E) are simulated with electromagnetic software. The dataset was collected in X-band (9.5-10.5GHz) across an azimuthal range of 0° to 60° (0.05° sampling step) and pitch angles from -15° to 15° (3° interval), encompassing four polarization models (HH, HV, VH, VV). This experimental setup produced 4×11×1201 range profiles per class, each consisting of 984 range bins. For each HRRP signal, SCs are limited through a peak detection scheme with calibrated parameters: a prominence value of 0.15, a minimum inter-peak spacing of 5 range cell, and an upper limit of 10 SCs. In the experiments, while forming the measured dataset, the "1" part of the tracks measured is selected.

*3.1.2 LLMs and Baselines:* Our experiments apply two latest (till Jun. 2025) LLMs from OpenAI, including GPT-4.1 and GPT-o4-mini for testing our strategy on different scales of LLMs. Baselines: Conventional machine learning approaches involve Support Vector Machines (SVM) trained on either HRRP amplitude features or SCs [13, 14] alongside Random Forest Models leveraging SCs [15]. We denote them as SVM-HRRP, SVM-SC, and RF-SC respectively. To ensure fair comparative analysis, FSL methods were excluded from the evaluation. This exclusion stems from their dependency on meta-training datasets for model adaptation.

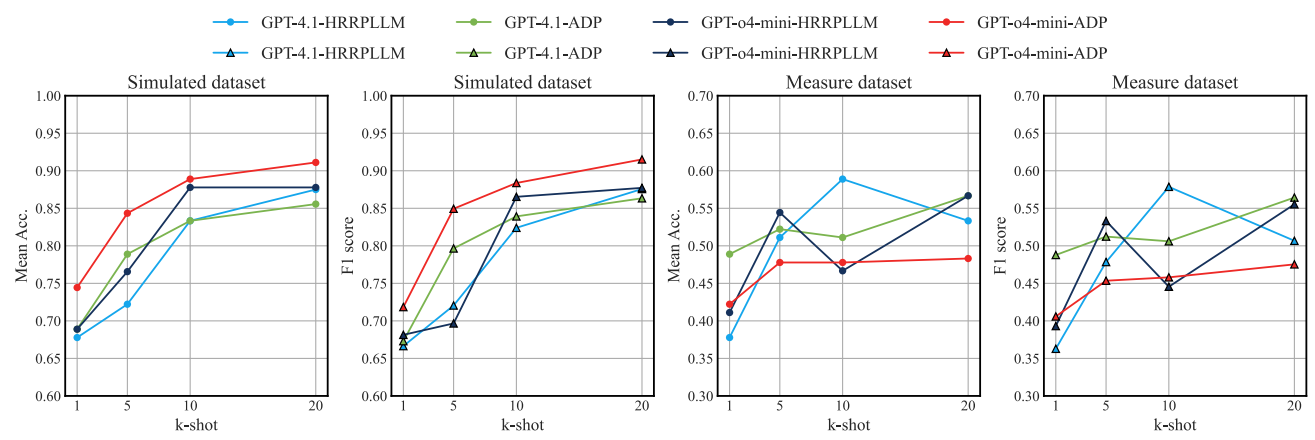

Figure 5 Impact of number of support shots (i.e., $K$ =1, 5, 10, 20) on HRRP ATR performance of the proposed method and HRRPLLM.

Table 1

**Results of Few-shot HRRP ATR Experiments**

The traditional ML methods and LLMs are calculated on simulated dataset. Each method is tested under four conditions, namely 1-shot, 5-shot, 10-shot and 20-shot. TF means Training-Free.

| Type | Model | K-shot | TF | Mean Acc. | F1. |
|---|---|---|---|---|---|
| Traditional ML | SVM-HRRP [13] | 1 | × | 62.22 | 54.44 |
| | | 5 | | 75.56 | 68.70 |
| | | 10 | | 90.00 | 86.67 |
| | | 20 | | 97.78 | 97.04 |
| | SVM-SC [14] | 1 | × | 42.22 | 33.15 |
| | | 5 | | 63.33 | 55.56 |
| | | 10 | | 72.22 | 65.37 |
| | | 20 | | 72.22 | 64.44 |
| | RF-SC [15] | 1 | × | 54.44 | 47.22 |
| | | 5 | | 71.11 | 65.93 |
| | | 10 | | 78.89 | 72.96 |
| | | 20 | | 87.78 | 83.70 |
| HRRPLLM [12] | GPT-4.1 | 1 | ✓ | 67.78 | 66.64 |
| | | 5 | | 72.22 | 72.05 |
| | | 10 | | 83.33 | 82.40 |
| | | 20 | | 87.50 | 87.57 |
| | GPT-o4-mini | 1 | ✓ | 68.89 | 68.14 |
| | | 5 | | 76.56 | 69.66 |
| | | 10 | | 87.78 | 86.53 |
| | | 20 | | 87.78 | 87.72 |
| **Proposed (Ours)** | GPT-4.1 | 1 | ✓ | 68.89 | 67.30 |
| | | 5 | | 78.89 | 79.65 |
| | | 10 | | 83.33 | 83.92 |
| | | 20 | | 85.56 | 86.33 |
| | GPT-o4-mini | 1 | ✓ | 74.44 | 71.84 |
| | | 5 | | 84.34 | 84.94 |
| | | 10 | | 88.89 | 88.37 |
| | | 20 | | 91.11 | 91.51 |

*3.2 Comparative results and analysis*

In this experiment, we compare the performance of the proposed ADP strategy with traditional ML baselines and HRRPLLM. In order to analyse impact of number of support shots on FSL performance, we set $K$ =1, $K$ =5, $K$ =10 and $K$ =20 for comparison. Tables 1-2 summarize these results. On the simulated HRRP dataset, LLMs perform well under all conditions, with accuracy monotonically increasing as K rises. Specifically, GPT-o4-mini achieves 88.89% 10-shot Mean Acc. and 88.37% F1., which outperforms both baselines and HRRPLLM by a maximum margin of 16.67% and 23.00%. GPT-4.1 achieves 78.89% 5-shot Mean Acc. and 83.92% F1., outperforming both baselines and HRRPLLM at largest 15.56% and 24.09%. On the measured HRRP dataset where significant aspect variance is more significant, proposed ADP strategy still shows strong performance compared with HRRPLLM. For 20-shot Mean Acc. and F1., GPT-4.1 outperforms it with maximum improvement of 3.34% and 5.76%. Results in Fig. 4 indicate that an increase in the number of support shots might promote performance. However, it should not be ignored that the performance of proposed HRRPLLM is worse than that of baselines under conditions of $K$ =10 and $K$ =20 in the measured HRRP dataset. One possible reason is that traditional ML methods support iterative training, while our method merely leverages LLMs for single-pass HRRP samples' judgment. When the number of support shots is small, traditional ML method performs poorly due to overfitting. However, when the sample size is bigger, the calculation of numerical decision boundary is more accurate than LLMs, thereby achieving better results.

Table 2
**Results of Few-shot HRRP ATR Experiments**
The traditional ML methods and LLMs are calculated on measured dataset. Each method is tested under four conditions, namely 1-shot, 5-shot, 10-shot and 20-shot. TF means Training-Free.

| Type | Model | K-shot | TF | Mean Acc. | F1. |
|---|---|---|---|---|---|
| Traditional ML | SVM-HRRP [13] | 1 | × | 53.33 | 45.93 |
| | | 5 | | 68.89 | 60.93 |
| | | 10 | | 77.78 | 71.67 |
| | | 20 | | 84.44 | 81.11 |
| | SVM-SC [14] | 1 | × | 50.00 | 41.30 |
| | | 5 | | 51.11 | 43.89 |
| | | 10 | | 52.22 | 43.15 |
| | | 20 | | 67.78 | 60.74 |
| | RF-SC [15] | 1 | × | 51.11 | 43.52 |
| | | 5 | | 58.89 | 51.67 |
| | | 10 | | 61.11 | 53.70 |
| | | 20 | | 78.89 | 72.59 |
| HRRPLLM [12] | GPT-4.1 | 1 | ✓ | 37.78 | 36.28 |
| | | 5 | | 51.11 | 47.84 |
| | | 10 | | 58.89 | 57.86 |
| | | 20 | | 53.33 | 50.67 |
| | GPT-o4-mini | 1 | ✓ | 41.11 | 39.31 |
| | | 5 | | 54.44 | 53.31 |
| | | 10 | | 46.67 | 44.55 |
| | | 20 | | 56.67 | 55.53 |
| **Proposed (Ours)** | GPT-4.1 | 1 | ✓ | 48.89 | 48.78 |
| | | 5 | | 52.22 | 51.23 |
| | | 10 | | 51.11 | 50.60 |
| | | 20 | | 56.67 | 56.43 |
| | GPT-o4-mini | 1 | ✓ | 42.22 | 40.59 |
| | | 5 | | 47.78 | 45.35 |
| | | 10 | | 47.78 | 45.80 |
| | | 20 | | 48.31 | 47.53 |

# 4 Conclusion

This paper proposes ADP strategy, a novel framework designed to address the critical challenge of aspect sensitivity in few-shot HRRP ATR using LLMs. We first identify a fundamental limitation in recently proposed HRRPLLM, that is its tendency to treat all support samples of a class as evidence for a monolithic prototype and overlook the significant intra-class variance caused by aspect sensitivity. To overcome this, ADP explicitly models aspect variance by constructing a distributed set of aspect-aware prototypes for each target class. These prototypes, derived from clustering support samples based on SC similarity, represent distinct aspect-dependent signatures, enabling the LLM to perform robust matching via ICL. Experiments on both simulated and measured HRRP datasets confirm the effectiveness of ADP. As results shown in Tables 1-2, ADP is more robust to aspect sensitivity. We also find that ADP successfully counters the accuracy drop observed in HRRPLLM on measured dataset as $K$ increases. This validates the practical utility of our proposed method. Our method reveals a new perspective for LLM-based HRRP ATR through precise designed features. Future work can be focused on the issue of efficient deployment on edge devices, more precise HRRP feature representation for LLMs and *etc*.